\documentclass{article}

\usepackage{PRIMEarxiv}

\usepackage[utf8]{inputenc} %
\usepackage[T1]{fontenc}    %
\usepackage{hyperref}       %
\usepackage{url}            %
\usepackage{booktabs}       %
\usepackage{amsfonts}       %
\usepackage{nicefrac}       %
\usepackage{microtype}      %
\usepackage{lipsum}
\usepackage{fancyhdr}       %
\usepackage{graphicx}       %
\graphicspath{{media/}}     %
\usepackage{float}
\usepackage{subcaption}
\usepackage{placeins}
\title{Measuring the Invisible: Evaluating the Impact of Public Funding on Open Source Software
}

\author{
  Domenech Burin, Laia \\
  Data Scientist \\
  Sovereign Tech Agency \\
  Berlin, Germany\\
  \texttt{laia@sovereign.tech} \\
}

\begin{document}
\maketitle

\begin{abstract}
Open Source Software (OSS) forms a critical layer of contemporary digital infrastructure, yet remains largely overlooked by the institutions and societies that depend on it. The 2021 log4j vulnerability exposed this dependence, prompting both private and public organisations to recognise OSS funding as an instrument for enhancing digital sovereignty and software supply chain security.  A pioneer model of public funding for OSS is the Sovereign Tech Agency (STA). With an extensive portfolio of complementary programmes, the agency invests in targeted software maintenance work, the people behind the code and the identification of critical vulnerabilities in digital infrastructure. Despite growing institutional interest, the causal impact of public funding on OSS project sustainability remains empirically unresolved. Existing literature is divided between econometric and socio-technical approaches with few attempts at causal identification. 

This work aims to bridge that divide by combining a Goal-Question-Metric framework with the Generalized Synthetic Control Method to estimate the causal effect of the Sovereign Tech Fund, the Agency's primary program, on OSS repository activity. Counterfactual trajectories are constructed from a matched donor pool of unfunded projects, enabling identification of what funded repositories would have looked like in the absence of intervention. The main results show that the funding has a significant positive effect on project velocity metrics: commits, pull requests —both merged and new ones—, and new issues. There is no significant effect on the number of releases, contributors or closed issues. This indicates that the STF funding mobilises existing development activity rather than expanding the contributor base or accelerating backlog resolution. These findings carry practical implications for the design of public OSS funding evaluation frameworks, where assessment metrics should be matched to programme objectives rather than applied uniformly across interventions.
\end{abstract}

\section{Introduction}

Open Source Software (OSS) is software whose source code is publicly available for inspection, use, redistribution and modification, which is often created in a decentralized manner and distributed for free. It is the cornerstone of the multiple components that shape the digital infrastructure with which we interact every day. Digital infrastructure can be thought of in the same way as physical infrastructure: access to water, roads to move on or electrical systems are used every day and and whose provision is guaranteed by States. Digital infrastructure works the same way - although we do not often see or think about the apps and software we use on a daily basis, they all rely upon free and public code to function (\cite{eghbal_roads_2016}). Over 96\% of codebases contain OSS (\cite{synopsys_open_source_security_and_risk_analysis_deep_2023}), and roughly 70-90\% of any software stack consists of OSS (\cite{open_source_security_foundation_open_2022}). Rather than being a form of community-based altruism, OSS is a critical cornerstone of our modern worlds’ economic, political and social systems. 

The significance of OSS embedded in digital infrastructure gained relevance in 2021, during the log4j vulnerability exposure which could have affected a wide range of applications: ranging from Twitter, to Hadoop, Spark to Steam and Minecraft, just to name a few (\cite{vaughan-nichols_log4shell_2021}). Incidents like this, along with the XZ utils backdoor discovered in 2024 (\cite{akamai_security_intelligence_group_xz_2024}) illustrate the consequences of societal oversight and under-maintenance of OSS projects. When a system relies on the unpaid labor of volunteer communities without offering reciprocal contributions, it becomes vulnerable to collapse if these communities can no longer sustain their efforts. This is the  classic “free rider” phenomena in economic commons. Resources are offered for free, and everybody (from individuals to technology giants) use them, but nobody is incentivized to contribute back (\cite{fremeaux_free-riding_2025}). 

Studies compare digital infrastructure to traditional commons, such as shared grazing lands, to illustrate how unchecked exploitation can lead to a "tragedy of the commons" (\cite{hoffmann_value_2024}). Communities involved in the maintenance and development of OSS projects often work on said projects in their spare time, juggling full time jobs or freelance work while they do so. This volunteer work dynamic comes with a culture that historically discourages discussions about monetary compensation (\cite{asparouhova_working_2020}). However, growing concerns around maintainer burnout (\cite{asparouhova_working_2020}), security vulnerabilities and widespread commercial freeriding on the labour of volunteers (\cite{geiger_labor_2021}) began to change this paradigm in OSS developer communities (\cite{salkever_open_2023}). 

Three models of funding for OSS projects can be distinguished: micro-donations, commercial funding and public funding (\cite{osborne_toolkit_2024}). Each model establishes a unique relationship between funders and projects, varies in the degree of community engagement, and produces different outcomes. For instance, individuals can support OSS developers through micro-donations via platforms like GitHub Sponsors (\cite{github_about_nodate}). These contributions are typically ad hoc and modest in scale, resulting in minimal interaction between funders and projects. Consequently, the long-term sustainability and impact of such funding remain uncertain (\cite{osborne_public-private_2024}).

Commercial funding has been the primary model of investment in OSS to the date through different formats. Direct funding methods include sponsorship of OSS consortia or foundations, which operate on a fee-paying membership model. It is also common for companies to sponsor OSS developers and to let their employees contribute to OSS projects during work hours, either as a job responsibility or as part of a voluntary initiative (\cite{butler_company_2021}). Companies can also participate in OSS development through support, collaboration, or hosting. The models described above fall under the "support" category. Collaboration, by contrast, involves multiple organisations sharing control over a project's intellectual property. In hosting arrangements, a single company exercises full control over governance and intellectual property, typically employing the maintainers and requiring contributors to sign contributor license agreements (CLAs), among other conditions  (\cite{osborne_characterising_2024}). 

Finally, States and the public sector have emerged as key funders of OSS with the recognition of funding as a policy to enhance digital sovereignty and security of software supply chains (\cite{bitkom_ev_leveraging_2026}; \cite{eurostack_initiative_foundation_ev_euro_2025}; \cite{komaitis_europes_2026}; \cite{singh_leveraging_2025}). A leading example of this kind of funding is the Sovereign Tech Agency (STA), which originated from the 2022 Sovereign Tech Fund (STF) program. Launched by the German Federal Ministry for Economic Affairs and Climate Action (BMWK) and hosted by SPRIND GmbH, the STF aims to enhance the resilience, security, and sustainability of critical open source digital infrastructure. The fund supports OSS projects through maintenance and development contracts starting at €50,000, with no upper limit, and project durations ranging from 6 to 24 months.

In November 2024, the STF transitioned into the STA, expanding its mission to develop a broader portfolio of initiatives (\cite{mucciacciaro_teuta_funding_2025}). According to its CEO, the STA is pioneering a new approach for governments to sustainably support open digital infrastructure while aligning with core values of security and sustainability (\cite{groh_maintaining_2025}). Beyond the original STF program, the STA now oversees three complementary initiatives: the Sovereign Tech Fellowship, which invests directly in developers; the Sovereign Tech Resilience program, focused on identifying and addressing critical vulnerabilities; and a third initiative designed to further strengthen OSS sustainability. Together, these programs create a comprehensive framework to address the multifaceted challenges of sustaining open source ecosystems.

In the aforementioned context, measuring the impact of the investment in OSS is an open and critical question relevant for both public and private actors. This question is an emerging field, open for different methodological approaches. This work aims to contribute to the field establishing a causal inference framework to measure the impact of public investment in OSS. The thesis will make two contributions:

\begin{enumerate}
  \item estimate the impact of the STA’s public funding for OSS, focusing on the STF program, and
  \item demonstrate that Generalized Synthetic Control is a viable  framework for evaluating the impact of OSS investments.
\end{enumerate}

The work is structured in the following way: in the next section, I will present  existing approaches to measuring OSS impact and existing gaps. Section \ref{sec:empirical_strategy} presents the empirical strategy used to address the question, the chosen model to address it, the data and the modeling specifications. Section \ref{sec:results} presents the results of Generalized Synthetic Controls for measuring impact of the Sovereign Tech Agency’s funding. Section \ref{sec:discussion} presents the discussion and limitations of this approach. Finally, section \ref{sec:conclusion} presents conclusions of this work. 

\clearpage

\section{Related Work}

Defining "impact" in the context of OSS is a prerequisite for designing an effective impact evaluation strategy. The literature shows two dominant approaches in this definition: economic and socio-technical. The first one aims to measure the indirect economic effects that the investment has on a broader industry or state level. This category includes valuation models such as the Constructive Cost Model (COCOMO) (Hoffmann et al., 2024), DevOps Research and Assessment (\cite{forsgren_science_2018}), ROI models (\cite{lawson_roi_2026}), or the underproduction risk model (\cite{champion_underproduction_2021}).  

However, these economic approaches face two major limitations. The first one is methodological: estimating the value of OSS is inherently challenging due to the heterogeneity of projects. OSS initiatives vary widely in size, age, organizational structure, and technical complexity. Most models rely on simplifying assumptions about project structures and code history, which often fail to capture the full intricacy of OSS ecosystems (Vargas, 2024). The second limitation is epistemological: over-reliance on technical metrics flatten socio-technical dynamics that characterise activity in OSS projects. This reductionist perspective overlooks the broader ecosystems of dependencies, contributors, and users, which are central to understanding the true impact of OSS.

The socio-technical approach addresses the second limitation by shifting focus to the social structure and health of OSS projects. Unlike purely economic models, this approach offers a more holistic lens for understanding value creation within OSS projects and their broader ecosystems. A central concept of this perspective is that impact can be assessed through the lens of project health: the sustainable availability of maintenance labor, whether from core maintainers or the wider community (Osborne, Sharratt, et al., 2024). OSS projects are a socio-technical interaction between code and community, and a "healthy" open source project is one that continues to produce, modify, and maintain code. Approaches following this model rely on qualitative methods, as they are the de-facto technique for identifying complex, contextual understanding of a phenomena without being constrained by assumptions of a quantitative model (Creswell \& Creswell, 2017; Leeming, 2018). An example of this approach are the Sovereign Tech Agency’s original impact reports: the Sovereign Tech Fellowship Evaluation (Wagner, 2025) or the Pilot Round Evaluation (SPRIN-D, 2023). However, the main limitation of these approaches is the possibility of generalization and measurability of the effects. 

Given these considerations, mixed methods approaches appear as the ideal approach for evaluating impact, as they combine both scalability and project complexity. In assessing the impact of investment on OSS projects, it is important to keep a qualitative dimension through interviews and firsthand testimonies from the maintainers and people involved in the communities to understand what is relevant for the projects (Osborne, Sharratt, et al., 2024). This approach acknowledges  that the definition of impact is a moving target: the effects of OSS investment depend on which dimensions of a project’s sustainability and health are being addressed, as well as the contextual knowledge provided by the communities involved. Thus, impact is not a fixed metric but a context-dependent outcome, shaped by the unique needs and insights of each OSS ecosystem.

\clearpage

\section{Empirical Strategy \& Data}
\label{sec:empirical_strategy}

This work aims to evaluate the effect of a treatment $T$ —specifically, funding from the STA through the STF program— on a set of outcomes $Y$. Building on the earlier discussion, the outcomes are defined based on the type of funded work undertaken by the projects. This way, the specific knowledge and targeted work defined by each community is maintained. 

To operationalize this, this work employs the Goal-Question-Metric (GQM) methodology (\cite{marciniak_goal_2002}). GQM starts with a qualitative assessment of the work packages outlined in the contracts between the Sovereign Tech Agency and the projects. It retroactively scopes these work packages to identify broad goals, refines those goals into specific questions, and ultimately translates them into measurable metrics \footnote{The descriptive approach—mapping defined work packages to quantifiable metrics—was developed during my work at the Sovereign Tech Agency. The novel contribution of this thesis is the causal leap, transitioning from descriptive metrics to a causal inference framework.}. These metrics are drawn from the dictionary developed by the CHAOSS (Community Health Analytics Open Source Software) group. CHAOSS is a Linux Foundation project focused on creating metrics, models, and software to assess open source community health globally. Together, GQM and CHAOSS form an integrated measurement chain: work packages from each funding contract are first mapped to goals and questions, and only then translated into the CHAOSS metrics defined in the following subsections, ensuring every metric is directly traceable to the work each project committed to deliver.

The assessed metrics can be divided into the following three dimensions:

\subsection*{Project Activity \& Engagement} 
These metrics capture the level of ongoing discussion, problem-solving, and community involvement: 
\begin{itemize} 
\item \textbf{New Issues:} Frequency of new issues created within a specified period. This metric can be interpeted in two ways. On one hand, a high volume indicates an active community discussing problems and proposing solutions, reflecting ongoing engagement and project vitality. On the other, it shows the increase of backlog and bugs in a given project. More recently, the rise of AI-generated noise in issue trackers has added a further source of noise to this metric (\cite{noauthor_metric_nodate-1}). 
\item \textbf{Closed Issues:} Number of issues closed during a certain period. A high volume signals a productive community actively resolving issues and completing work. (\cite{noauthor_metric_nodate}). 
\end{itemize} 
\subsection*{Development \& Code Contribution} 
These metrics assess the health and sustainability of the codebase, as well as the frequency and acceptance of contributions: 
\begin{itemize} 
\item 
\textbf{Commits:} Number of atomic changes to the source code over a specific period. High commit volume indicates active development, maintenance, and responsiveness to bugs or feature requests. (\cite{noauthor_metric_nodate-2}). 
\item \textbf{New Change Requests:} Number of proposals for code modifications submitted for review. This metric reflects coding activity and contributor engagement with the codebase.  (\cite{noauthor_metric_nodate-6}).
\item \textbf{Merged Change Requests:} Number of accepted and merged change requests. This serves as a proxy for project vitality and the volume of successful contributions. (\cite{noauthor_metric_nodate-3}). 
\end{itemize} 

\subsection*{Community \& Dynamics} This metric evaluates the breadth and depth of contributor involvement: \begin{itemize} 
\item 
\textbf{Contributors:} Number of individuals contributing code, documentation, or other resources. A higher number of contributors is associated with project health, longevity, and access to necessary resources. (\cite{noauthor_metric_nodate-4}).

\end{itemize} 
\subsection*{Project Maturity \& Release Activity} 
This metric reflects the project’s ability to deliver updates and respond to user needs: 
\begin{itemize} 
\item \textbf{Releases:} Frequency of software/artifact releases over time. Consistent release frequency may indicate a stable or mature project, while rapid iteration suggests responsiveness to user needs.  (\cite{noauthor_metric_nodate-5}). 
\end{itemize}

These dimensions are critical for assessing both the technical and social health of open source projects, as well as their capacity for innovation and sustained development. However, it is worth noting that these metrics show only one side of the story. While an increase in commits indicates a rise in activity and development, it also means that as projects expand, the demands on contributors increase. A higher number of pull requests and merged requests indicate an active community contributing to improving the software, but can also mean that said software has many issues to fix. Therefore, these metrics have to be taken contextually alongside a qualitative interpretation to effectively assess sustainability and health. This work aims to enhance the quantitative approach and provide a causal framework, but it does not intend to fully replace the assessment of the impact of open-source software funding. These metrics are a compass: they indicate the direction, but not the destination. 

While the project portfolio of the STF currently contains 187 projects, this work will focus on four:

\begin{itemize}
    \item \textbf{Python Package Index (PyPI):} Python is one of the most widely used programming languages for developing websites, software, and machine learning and AI models in today's development landscape (\cite{python_software_foundation_python_2024}). Package management in Python is handled through the pip ecosystem, which hosts its packages on PyPI. With over 1,046,786 users and 794,245 projects, it stands out as one of the primary package repositories for Python (\cite{python_package_index_pypi_nodate}). The STF invested on critical engineering, maintenance and support to PyPI as well as some of its highly used and critical packages that developers rely upon to secure their applications. This work focused primarily on implementing new features to strengthen the overall ecosystem.
    \item \textbf{curl:} curl is an open source tool used in command lines or scripts to transfer data. The library behind it, libcurl, is one of the world’s most used libraries. It is widely used from countless software applications to cars, television sets, routers, printers, audio equipment, mobile phones, tablets, medical devices, and more. It has over twenty billion installations (\cite{curl_curl_nodate}). The funding of the STF targeted maintenance via addressing technical backlog: resolving 122 known bugs through systematic classification and fixes, while also introducing new options to support newer versions of the HTTP protocol.
    \item \textbf{Fortran:} Fortran is a widely used programming language that has been designed from the ground up for computationally intensive applications in science and engineering. Today, it is the standard in domains that adopted computation early–science and engineering. These include numerical weather and ocean prediction, computational fluid dynamics, applied math, statistics, and finance. It is particularly well-established in scientific communities due to its reliability (\cite{fortranlang_fortran_nodate}). STF investment focused on modernizing the Fortran ecosystem through improvements to the compiler and the package manager, ensuring continued usability and relevance for contemporary scientific computing.
    \item \textbf{RubyGems:} RubyGems and Bundler provide library and software package management for the Ruby programming language, included with every copy of Ruby in the language standard library. They are indispensable tool for using Ruby, and many applications rely on their regular maintenance and improvement, as well as their ongoing usability \cite{rubygems_bundler_nodate}.
    RubyGems.org hosts publicly available software packages (called “gems”) which enables developers to quickly and securely integrate more functionalities into their projects, while Bundler downloads and installs the exact gems and versions needed for a given project. The STF funding was used to strengthen community maintenance, improving the reliability, security and scalability of RubyGems.org and improve maintainer quality of life by reducing repetitive work and risk (via automation, clearer tools, and safer defaults), giving maintainers and contributors better visibility, workflows, and support systems to collaborate effectively.
\end{itemize}

 The decision to select these was based on two main focuses: on one hand, these are the oldest projects, being implemented on the pilot round of the STF. This means that there are several post-funding time periods which make them feasible for evaluation. On the other hand, these four projects addressed a variety of types of maintenance activities, making it possible to evaluate different metrics across targeted work. The implications of this heterogeneity are further discussed in Section~\ref{sec:discussion}.

Data was collected via \href{https://ecosyste.ms/}{\texttt{ecosyste.ms}}, an open and transparent tool that indexes billions of events to create one of the world’s most comprehensive datasets on open source production and use. Ecosyste.ms hosts data for over 287 million repositories. The data collection pipeline is documented in the \href{https://github.com/ldmnch/ecosystems_data_collection}{\texttt{ecosystems\_data\_collection}} GitHub repository, developed as part of the main deliverable of this thesis.

While the GQM approach provides a descriptive analysis of these projects, and a way to map overall project goals to specific metrics, this work seeks to advance the methodology by developing a semi-experimental causal framework. The goal is to move beyond description and empirically establish how STF funding differentially affected the long-term sustainability of the treated units. The main idea behind potential outcomes (\cite{rubin_causal_2005}) is that for each unit $i$ there are two potential outcomes: $Y_i(1)$ if treated and $Y_i(0)$ if not treated. However, the fundamental problem is that we can never observe both for the same unit. One potential outcome is always missing. A naive comparison between treated projects (funded STF projects) with untreated (any other OSS project, not funded by the STF) can be biased. The groups may differ systematically due to confounders. This has been previously described in the challenges in causal methods for OSS impact evaluation: identifying comparable units for the control group. Constructing a reliable counterfactual is a challenge, as software projects have multiple covariates across which they might vary: such as size, lifecycle stage, community composition, adoption and usage. 

The Synthetic Control Method (SCM) (\cite{abadie_synthetic_2010}) addresses this by constructing counterfactual trajectories for treated units from non-experimental data. Rather than relying on a single comparison unit, SCM forms a weighted combination of untreated units (the "donor pool") chosen to closely approximate the treated unit's pre-intervention trajectory. However, the validity of this approach depends critically on donor pool composition: if the pool contains units that are structurally dissimilar from the treated unit, the resulting synthetic control may be unreliable, leading to erroneous causal conclusions (\cite{abadie_comparative_2015}).

To build a reliable control donor pool, this study focused on the core criterion for Sovereign Tech Agency funding: critical digital infrastructure. This means that these projects are open software components that are vital to the development of other software or enable digital networking (\cite{sovereign_tech_agency_sovereign_2026}). To identify such projects, I worked with the open dataset of the OpenSSF Scorecard, which contains a scan of the 1 million most critical open source projects judged by their direct dependencies. Using this dataset as a baseline, package metadata (dependent packages and level of usage) was scraped from these repositories using Ecosyste.ms. 

With this information, propensity score matching (\cite{rosenbaum_central_1983}) was used to produce the donor pool. Propensity score matching (PSM) is a statistical technique designed to reduce selection bias in observational studies by simulating the conditions of a randomized experiment. The method works by estimating the probability that a unit (in this case, an open source repository) receives treatment (STF funding), conditional on observed pre-treatment characteristics. These characteristics, or covariates, are chosen to reflect the criticality and comparability of projects, including metrics such as the number of dependent repositories, dependent packages, stargazers, forks, and ownership status (public vs. private).

\begin{itemize}
    \item Dependent repositories: count of repositories that depend on the packages developed in a repository.
    \item Dependent packages: count of packages that depend on the packages developed in a repository.
    \item Stargazers: count of stargazers of a repository.
    \item Forks: count of forks of a repository.
    \item Average ranking: average score for the ranking of the packages published within the project, based on usage metrics like downloads, dependent packages and repositories, and compared with other packages from the same package manager registry. 
    \item Private ownership: if the project is open, but owned by a private company (e.g.: open AWS software). 
\end{itemize}

The logic behind this is that repositories with similar levels of dependents, stargazers and forks will have similar levels of criticality and therefore could potentially be funded by the STA (\cite{nesbitt_how_2025}). Criticality is defined as the influence and importance of a project, assessed based on its impact on users, which includes factors such as its influence, security exposure, and usage intensity \cite{open_source_security_foundation_criticality_nodate}. By matching treated units (funded repositories) to control units (unfunded repositories) with similar propensity scores, PSM helps ensure that the comparison is made between projects that are as alike as possible in terms of their baseline characteristics. This approach mitigates confounding and strengthens the validity of causal inferences about the impact of STF funding on project outcomes.

A 1-to-1 nearest-neighbour propensity score matching was performed at the package level using a caliper of 0.0001, returning 62 distinct control repos matched to the 12 treated units. Table ~\ref{tab:covariate_balance} shows the covariate balance of the means before and after the PSM. PSM reduced the mean absolute standardized mean difference (SMD) from 0.466 to 0.119 (a 74\% reduction) with 3 of 6 covariates achieving the conventional |SMD| < 0.10 threshold (\cite{austin_balance_2009}). For this specific case, I am not interested in achieving perfect balance across all controls. The main purpose of this step is to narrow the donor pool to structurally comparable controls (\cite{ho_matching_2007}). Residual imbalance on popularity metrics (\texttt{dependent\_repos\_count}, \texttt{stargazers\_count}, \texttt{forks\_count}) is subsequently addressed by the Generalized Synthetic Control estimator through its interactive fixed effects structure (\cite{xu_generalized_2017}), described in the following section.

\begin{table}[ht]
\centering
\caption{Covariate Balance Table (Propensity Score Matching via Pymatch)}
\label{tab:covariate_balance}
\resizebox{\textwidth}{!}
{%
\begin{tabular}{lrrrrrrr}
\toprule
\textbf{Covariate} & \textbf{Mean Treated} & \textbf{Mean Control} & \textbf{SMD} & \textbf{Mean Treated} & \textbf{Mean Control} & \textbf{SMD} \\
 & \textbf{(Before)} & \textbf{(Before)} & \textbf{(Before)} & \textbf{(After)} & \textbf{(After)} & \textbf{(After)} & \\
\midrule
\texttt{Dependent repositories}     & 30.198 & 15.955 & 0.824 & 28.902 & 24.550 & 0.229 \\
\texttt{Dependent packages}  & 14.836 & 18.352 & 0.244 & 14.768 & 14.536 & 0.019 \\
\texttt{Stargazers}           &  3.264 &  3.972 & 0.129 &  3.444 &  4.554 & 0.183 \\
\texttt{Forks}                &  2.641 &  4.146 & 0.304 &  2.815 &  3.980 & 0.242 \\
\texttt{Average ranking}            & 21.516 & 12.082 & 0.915 & 20.728 & 20.278 & 0.038 \\
\texttt{Private ownership}              &  0.000 &  0.068 & 0.382 &  0.000 &  0.000 & 0.000 \\
\bottomrule
\end{tabular}%
}
\end{table}

One repository can host multiple packages, so the final sample size is as expressed in Table~\ref{tab:sample_size}.

\begin{table}[h] 
\centering 
\caption{Sample Size by Repository Group} 
\label{tab:sample_size} 
\begin{tabular}{ccc}
\hline 
\texttt{Repository Group ID} & \texttt{Repository Group} &  \texttt{N} \\ 
\hline 101 & Python Package Index & 5 \\ 
102 & RubyGems & 3 \\ 
104 & Fortran & 3 \\ 
105 & curl & 1 \\ 
204 & Control Group & 62 \\ 
\hline 
\end{tabular} 
\end{table} 

For each repository, historical data was collected from the ecosyste.ms API (commits, pull requests, issues, releases). The data was filtered from 2015 onwards and grouped by quarters, resulting in 54 pre-treatment periods per repository. This provides a substantial panel dataset for ASCM and GSCM models. Treatment timing was defined as the quarter in which the STF contract was signed for each repository. As treatment onset is staggered across projects, each unit's pre- and post-treatment periods are defined relative to its individual funding date, which is directly accommodated by the models' staggered adoption design.

The experimental dataset consists of outcome variables per quarter and repository. Due to high sparsity, all outcomes were log-transformed to normalize distributions and stabilize variance. Outliers were trimmed by removing repositories with standard deviations exceeding the 95th percentile, ensuring stable donor trajectories and reducing distortion in the synthetic counterfactual.

\subsection{Methodology}

Differences-in-differences is one of the most commonly used empirical designs to evaluate the impact on different programmes. Its key assumptions are the stable unit treatment value, lack of anticipation effect for the treated units, and parallel trends - meaning that in the absence of treatment, both control and treated units would have followed parallel paths on the evolution of the outcome variable. However, these three assumptions often do not hold in non-experimental settings. The parallel trends assumption is particularly challenging: it cannot be directly tested, and while researchers often examine pre-treatment trends in the outcome variable as a form of descriptive evidence, this approach does not constitute a rigorous validation.

In semi-experimental settings, it is challenging to find a single unexposed unit that approximates the relevant characteristics of the units exposed to treatment and therefore have a parallel development. However, a weighted combination of units can provide a good comparison for the treated one: this is the premise behind the synthetic control method. Developed by \cite{abadie_synthetic_2010}, this method averages multiple control units by weighting them as representations of the counterfactual for one treated unit. Each control unit is restricted to a weight to be positive and sum to one, making explicit the relative contribution of each control unit to the counterfactual of interest. 

The synthetic control method has become the main quantitative approach for causal inference in case studies. Providing a systematic way to choose comparison units in comparative case studies, it opens the door for precise quantitative inference in small-sample comparative studies, without excluding the application of qualitative approaches (\cite{abadie_comparative_2015}). In this sense, the method provides a bridge between qualitative and quantitative approaches in empirical research, making it the ideal technique to approach the issue of measuring impact in OSS. Its ability to identify comparable projects based on diverse characteristics further underscores its value for this type of empirical analysis.

However, the synthetic control method is suitable only for the case of one treated unit. The setup for the synthetic control method is that, from a sample of $J + 1$ units, there is one $j = 1$ treated unit and units $j = 2$ to $j = J + 1$ are the potential comparisons. Given the size of the STA’s project portfolio, this sole method is not enough to achieve the goal of finding a generalizable method to evaluate impact across groups of projects. This experiment works with four projects that add up to twelve repositories (treated units) with different treatment adoption times and 62 units for the donor pool based on the propensity score matching approach. Therefore, for the goal of this thesis the initial approach explored two variations of the synthetic control method that overcome this limitation. 

The Augmented Synthetic Control Method (ASCM) (\cite{ben-michael_augmented_2021}) is a doubly robust approach that combines an outcome model to estimate and correct for bias arising from imperfect pre-treatment fit. It augments the original synthetic control method by incorporating a ridge regression estimator, which can be interpreted as a weighted average of control unit outcomes. This regularization improves the pre-treatment fit of the estimated weights. While the synthetic control method restricts the sum of the weights to positive values and sum to one, Augmented Synthetic Controls allow for negative weights on some units. The method has been applied to policy evaluation and clinical research settings (\cite{krajewski_augmented_2024}; \cite{zhao_application_2025})

The second approach is the Generalized Synthetic Control Method (GSCM) (\cite{xu_generalized_2017}). This model relaxes the assumption of parallel trends and the single-unit limitation. It unifies the synthetic control method with assumptions and basic method of synthetic controls with an interactive fixed effects model (\cite{bai_panel_2009}) enabling semiparametric modelling of unobserved time-varying confounders within a unified framework. The estimation proceeds in three steps: the interactive fixed effects model is fitted on the control units to obtain a set of latent factors, the factor loadings for each treated unit are estimated by projecting pre-treatment outcomes onto the space spanned by those factors and treated counterfactuals are imputed using the estimated factors and loadings. By drawing on all available control observations, GSCM improves estimation efficiency relative to standard synthetic control approaches and produces frequentist uncertainty estimates including standard errors and confidence intervals. 

In the next section, I will describe the modeling strategy applied to both methods and results achieve to define which of these show a better fit for the data.

\subsection{Model Selection and Specifications}

Treatment and pre-treatment periods are defined based on the date each repository receives funding from the Sovereign Tech Agency, resulting in staggered treatment adoption across units. The staggered design is accommodated by estimating treatment effects relative to each unit's individual treatment date.

Both models incorporate latent factors to address unobserved confounders. For GSCM, the optimal number of factors is determined via internal cross-validation. For ASCM, which lacks automated factor selection, a manual search across one to ten factors is conducted, selecting the specification that minimises the Panel Criteria  (\cite{bai_determining_2002}). Table~\ref{tab:model_factors} summarises the optimal factor count for each model and outcome. While the factors itself are not interpretable, as they represent an abstract dimension of unobserved confounding structure, both of the models  converge on similar factor counts across outcomes, diverging by no more than three in any case. This consistency suggests that both estimators identify a broadly comparable latent structure in the data.

Pre-treatment fit assessments (Appendix Figures~\ref{fig:gsynth_att} and~\ref{fig:agusynth_att}) show that GSCM yields tighter confidence interval bands across all outcomes. Beyond this empirical advantage, GSCM's interactive fixed effects structure is more directly suited to modelling the unobserved time-varying confounders that characterise OSS repository dynamics. Based on superior pre-treatment fit and stronger theoretical alignment with the data-generating process, GSCM is selected as the primary estimator for all subsequent results, with ASCM reported as a robustness check.

\clearpage

\section{Results}
\label{sec:results}

This section presents the estimated average treatment effects of Sovereign Tech Agency funding on the selected repository-level outcomes, as estimated by the Generalized Synthetic Control Method. Counterfactual trajectories are constructed using the pre-treatment period to project what repository activity would have looked like in the absence of funding. 

\begin{table}[H]
\centering
\caption{Average Treatment Effect on the Treated (ATT) by Outcome}
\label{tab:att_results}
\begin{tabular}{lcc}
\toprule
\textbf{Outcome} & \textbf{ATT (log)} & \textbf{p-value} \\
\midrule
Commits       & \phantom{-}0.8912 & 0.0064\hspace{1em}**  \\
Contributors  & \phantom{-}0.0283 & 0.9346\hspace{1em}\phantom{***} \\
Releases      & -0.6476           & 0.0585\hspace{1em}\phantom{***} \\
Closed Issues & \phantom{-}0.8178 & 0.1559\hspace{1em}\phantom{***} \\
New Issues    & \phantom{-}1.2501 & 0.0001\hspace{1em}*** \\
Merged CRs    & \phantom{-}1.0133 & 0.0397\hspace{1em} **   \\
New CRs       & \phantom{-}0.8661 & 0.0380\hspace{1em} **   \\
\bottomrule
\end{tabular}
\caption*{\footnotesize\textit{Note: *** p $<$ 0.01, ** p $<$ 0.05, * p $<$ 0.10.}}
\end{table}

Table \ref{tab:att_results}  presents the ATT of each outcome with their respective significant level at 1\%, 5\% and 10\%. As described in the modeling specifications, all the variables were set to a log scale and the uncertainty estimates are constructed on a frequentist sample of 1,000 repetitions. The findings show that the STF intervention produces statistically significant effects across: the overall number of change requests, the number of merged change requests, the number of new issues and the number of commits. Contributors, releases and closed issues do not show statistical significance of an effect. 

The estimated ATT show positive outcomes for the metrics related to the project velocity, which captures the pace of development activity. The ATT estimates indicate that funded projects experience, on average, a 143.8\% increase in commits, 175.5\% in merged change requests, 137.8\% in new change requests, and 249\% in new issues. These percentage changes are derived by applying the inverse log transformation ($(e^{\hat{\delta}} - 1) \times 100$) and should be interpreted as multiplicative effects relative to the counterfactual trajectory. Figure \ref{fig:gsynth_att} shows the evolution of the treatment effect over time with the respective confidence interval for each time period.  The wide confidence intervals across all statistically significant metrics reflect considerable heterogeneity in the ATT estimates among projects. These outcomes should therefore be treated as directional signals of the positive effect of funding rather than precise estimates.

Table \ref{tab:model_factors} reports the optimal number of factors for each outcome. New issues require the highest number of factors, suggesting a more complex latent confounder structure driving its dynamics. These factors and factor loadings are not directly interpretable as they are linear transformations of the true underlying factors. However, the consistently high factor counts across outcomes indicate that latent confounders play a meaningful role in shaping repository activity. 

Figures \ref{fig:gsynth_counterfactuals} and \ref{fig:gsynth_att} together illustrate the adequacy of the pretreatment fit for each outcome. The ATT plots in Figure \ref{fig:gsynth_att} show distinct spikes at time periods $-11$ and $-4$ across all models. For statistically significant outcomes, confidence intervals remain tight and centered around zero throughout the pretreatment period, with no evidence of anticipation effects. Releases and contributors diverge from this pattern: releases display persistent spikes exceeding zero with wide confidence intervals, while contributors exhibit broader uncertainty than the remaining metrics. Both of these suggest weaker pretreatment fit and greater underlying volatility.

The main insight these results indicate is that the STF produces an overall stimulus on new activity and engagement (submission and merging of new change requests, committing). The increase of issues admits two interpretations: on one hand, it can signal community engagement, as new developments prompt users to report bugs and request features. Alternatively, it reflects an expanding backlog that is not matched by a corresponding increase in closed issues, suggesting that projects could struggle to absorb the additional activity. Lastly, the absence of an effect on the number of contributors indicates that this funding program does not translate into measurable community growth. The increased activity appears to be driven by existing contributors rather than by an expansion of the contributor base.

\clearpage

\section{Discussion}
\label{sec:discussion}

The results presented above demonstrate a substantial and positive impact of the Sovereign Tech Agency’s public funding on OSS repository activity. However, interpreting these findings requires careful consideration of the study’s limitations. The following discussion on the study’s findings focus on three interconnected sources of uncertainty: control group construction, outcome metric selection, and model fit. 

\textbf{Control Group Selection.} As discussed in the Empirical Setting \& Data section, the heterogeneity of OSS projects poses a fundamental challenge for causal identification. To address this challenge, this work employed comparative case studies to bridge the qualitative-quantitative divide and mitigate potential confounders and bias. The combination of propensity score matching and the Generalized Synthetic Control Method mitigates observable confounding by selecting donor units that reflect the Sovereign Tech Agency's funding eligibility criteria and capturing project criticality through matched covariates. Nevertheless, several dimensions relevant to comparability remain unaddressed in the current specification, including project vulnerability, lifecycle stage, and salary structures or cost heterogeneity across regions and organisations. 

The most suitable solution would be to construct the donor pool from projects shortlisted by the Sovereign Tech Agency: those close enough to receive funding but excluded due to financial constraints. Such a pool would offer a stronger basis for causal comparison. This approach was not feasible for the present study due to ethical constraints, as these projects did not consent to data use for research purposes, and represents an important direction for future work should access be negotiated.

Future work could partially address the remaining gaps through additional proxies, such as the ratio of core maintainers to lines of code as an indicator of structural fragility, and repository creation date as a proxy for lifecycle stage. The reliability of such proxies depends heavily on data quality and completeness, which remains limited for the repositories in this sample. It should also be noted that some of these dimensions are partially captured by the existing matching covariates: projects with high dependent counts, forks, and clones are likely to share similar lifecycle stages and organisational scales, providing an implicit though imperfect control. Vulnerability and lifecycle stage nonetheless remain incompletely addressed, and future specifications should consider incorporating these dimensions explicitly to strengthen donor pool validity.

\textbf{Outcome Metrics.} The metrics selected to evaluate funding impact are project-level indicators of broad repository activity. A substantive limitation of this approach is that Sovereign Tech Agency contracts specify different work packages across projects: some target backlog resolution while others focus on new feature development or security hardening. This divergence in funding scope represents a potential threat to internal validity, as the same outcome metric may capture fundamentally different underlying processes depending on the project. Generalised metrics of the kind used here capture broad trends in repository activity but may obscure treatment effect heterogeneity driven by differences in funding intent.

A complementary analytical approach would be to stratify projects by contract type and construct outcome-specific indices composed of more granular metrics tailored to each funding category. This would allow for a more precise evaluation of treatment effects within comparable groups of projects, reducing the confounding influence of divergent work package objectives. Such stratification is beyond the scope of the present study, which prioritises the estimation of generalised funding effects across the full sample. A challenge that arises from this approach is that disaggregating by funding objective (security work versus backlog resolution, for instance) complicates the construction of valid control groups. Nonetheless, this represents a productive direction for future research, and subsequent work at the STA will seek to operationalise this framework.

\textbf{Generalized Synthetic Control Method Fit.} Pre-treatment fit, as measured by the mean squared prediction error reported in Table~\ref{tab:model_factors}, is acceptable for most outcomes. Because all outcomes are specified on a log scale, MSPE values are not directly comparable to raw-unit benchmarks and should be interpreted relative to the variance of the outcome series rather than in absolute terms. Commits and Contributors exhibit comparatively higher MSPE values (0.71 and 0.52 respectively), consistent with the wider confidence intervals and pre-treatment volatility visible in Figure~\ref{fig:gsynth_att}. These outcomes are inherently noisier activity-based metrics and are more sensitive to idiosyncratic repository events unrelated to funding. The limited size of the donor pool is a contributing factor, as a larger and more diverse pool of control units would provide GSCM with a richer basis for constructing counterfactual trajectories and would be expected to improve pre-treatment fit and reduce ATT uncertainty for these outcomes. Notwithstanding these limitations, the tighter fit achieved for the remaining outcomes and the consistency of the direction of effects across all metrics support the overall conclusion that Sovereign Tech Agency funding had a positive impact on repository activity.
\clearpage
\section{Conclusion}
\label{sec:conclusion}
This work investigated the causal effect of STA's programme STF across a subset of metrics related to sustainability of OSS projects, employing a Generalized Synthetic Control Method to estimate the Average Treatment Effect on the Treated across seven repository-level outcomes. The results indicate that STF funding produces statistically significant and positive effects on four dimensions of project activity: commits, new change requests, merged change requests and new issues. By contrast, no statistically significant effects were detected for releases, closed issues, or contributors. Taken together, these findings suggest that STF funding stimulates new activity and engagement within funded repositories. However, they do not not find evidence that it accelerates the resolution of existing backlog, development of new releases or the increase of a contributor base. 

These results carry meaningful implications for both the evaluation and design of public investment in OSS infrastructure. The asymmetry between activity-stimulating and work-resolving effects, alongside the absence of contributor growth, point to a nuanced picture of what funding achieves: it mobilises targeted development work across projects, but shows limited evidence of expanding the contributor base or reduce the open backlog. These patterns are consistent with a scenario in which the funding stimulates development activity, but money alone does not resolve the binding constraint of maintainer capacity that can help sustain projects in the long-term. 

However, these findings are to be read also in the context of the STA’s larger funding portfolio. As described in the introduction, two other funding programmes of the STA address specifically the people behind the code and the resolve of critical issues that threat against digital infrastructure. This raises the possibility that STF funding shouldn’t necessarily be expected to show contributor growth or issue closing, as those are the Fellowship's or Resilience’s remit. If so, the absence of effects on these dimensions could reflect programme design rather than a fundamental limitation of public funding for OSS infrastructure. 

From a policy perspective, this has practical relevance for funding bodies and designing evaluation frameworks. Given the variation in objectives across programmes and the individual work packages within STF contracts, assessing interventions using mismatched indicators risks misrepresenting their impact. Evaluating a fellowship on commit velocity, or a fund contract on contributor growth, risks misrepresenting what an intervention was designed to achieve. Future work should address the construction of a comprehensive typology of metrics aligned to programme types and work contracts to evaluate the impact. 

The methodological contribution of this work also lies in demonstrating that the GSCM, combined with propensity score matching for donor pool construction, offers a promising quasi-experimental framework for evaluating OSS funding interventions: a domain where causal identification has been limited by project heterogeneity and the absence of randomised assignment.

In sum, this study contributes to a growing body of work seeking to rigorously evaluate public investment in Open Source infrastructure, a domain that has historically relied more on descriptive accounts than on causal evidence. By estimating the effects of STA funding using a quasi-experimental framework, it provides empirical insight into how targeted financial interventions shape repository-level dynamics. The findings highlight both the potential and the limitations of such interventions: while funding can effectively stimulate development activity and engagement, it does not, on its own, resolve deeper structural constraints related to maintainer capacity, backlog management, or community expansion. 

Future work should seek to address the limitations identified in this study, including the restricted size of the donor pool, the incomplete covariate specification in the matching stage, and the absence of contract-type stratification in the outcome metrics which leads to heterogeneous effects. In particular, expanding the pool of comparable control repositories and incorporating richer project-level covariates would improve the credibility of counterfactual estimation. Collaboration with the STA to broaden access to data on both funded and non-funded projects, as well as to develop outcome indices tailored to specific work package categories, would substantially strengthen the precision, interpretability, and generalisability of subsequent analyses.

These results point to the importance of understanding open source sustainability as a multidimensional phenomenon, in which financial resources interact with organisational, social, and technical factors. As OSS continues to function as critical digital infrastructure, the design and evaluation of funding mechanisms will require increasingly nuanced and evidence-based approaches. Strengthening collaboration between funding bodies, researchers, and OSS communities-particularly in improving data availability and aligning evaluation metrics with program objectives—will be essential for developing more effective and accountable models of public support in this space.

\FloatBarrier
\newpage
\section{Annex}
\vspace{-0.5em}
\subsection{Figures}
\vspace{-1em}
\begin{figure}[!ht]    
\centering
    \begin{subfigure}{0.48\textwidth}
        \includegraphics[width=\textwidth]{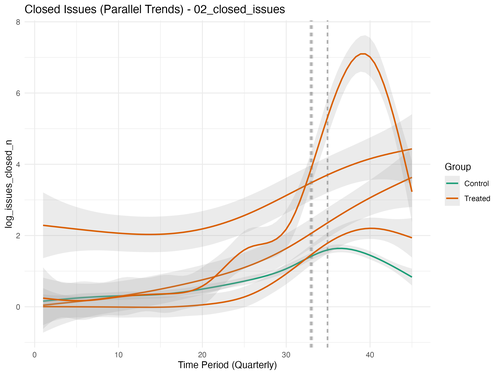}
        \caption{Closed Issues}
    \end{subfigure}
    \hfill
    \begin{subfigure}{0.48\textwidth}
        \includegraphics[width=\textwidth]{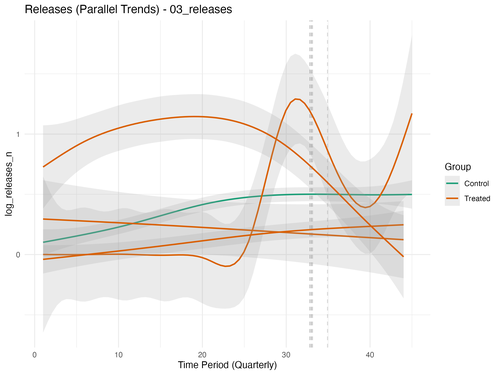}
        \caption{Releases}
    \end{subfigure}

    \vspace{0.5cm}

    \begin{subfigure}{0.48\textwidth}
        \includegraphics[width=\textwidth]{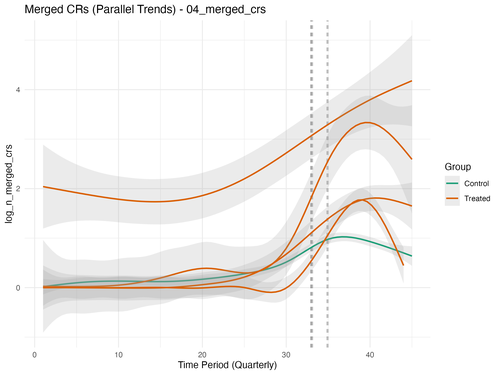}
        \caption{Merged CRs}
    \end{subfigure}
    \hfill
    \begin{subfigure}{0.48\textwidth}
        \includegraphics[width=\textwidth]{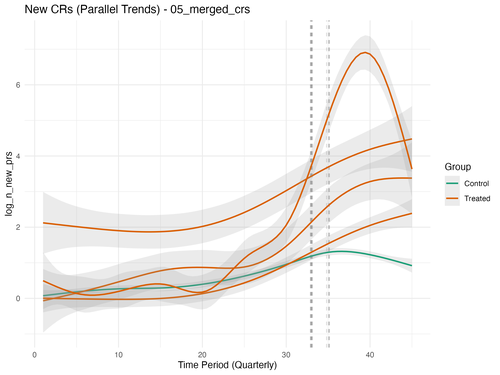}
        \caption{New CRs}
    \end{subfigure}
    \vspace{0.5cm}
    \begin{subfigure}{0.48\textwidth}
        \includegraphics[width=\textwidth]{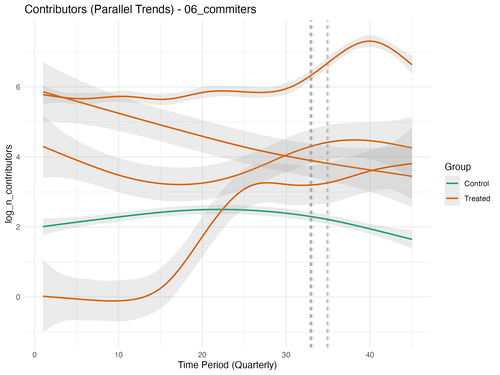}
        \caption{Contributors}
    \end{subfigure}
    \hfill
    \begin{subfigure}{0.48\textwidth}
        \includegraphics[width=\textwidth]{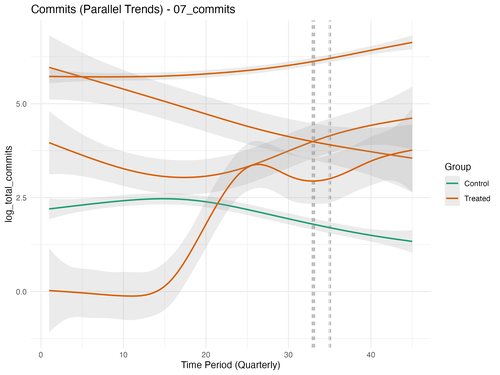}
        \caption{Commits}
    \end{subfigure}
    \caption{Trend of Outcomes Over Time (1 of 2)}

\end{figure}
\begin{figure}[htbp]
    \ContinuedFloat 
    \centering

    \begin{subfigure}{0.48\textwidth}
        \includegraphics[width=\textwidth]{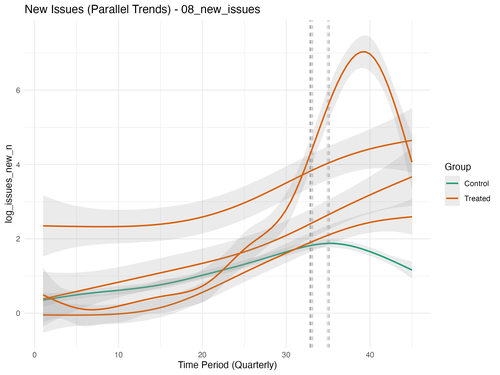}
        \caption{GSCM Counterfactuals: New Issues}
    \end{subfigure}
    \caption{Trend of Outcomes Over Time (2 of 2)}
    \label{fig:paralell_trends_test}
\end{figure}

\begin{figure}[htbp]
    \centering
    \begin{subfigure}{0.48\textwidth}
        \includegraphics[width=\textwidth]{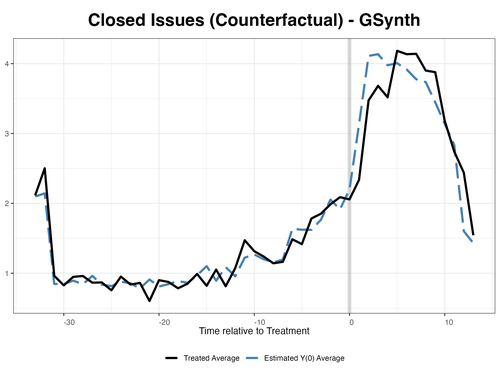}
        \caption{Closed Issues}
    \end{subfigure}
    \hfill
    \begin{subfigure}{0.48\textwidth}
        \includegraphics[width=\textwidth]{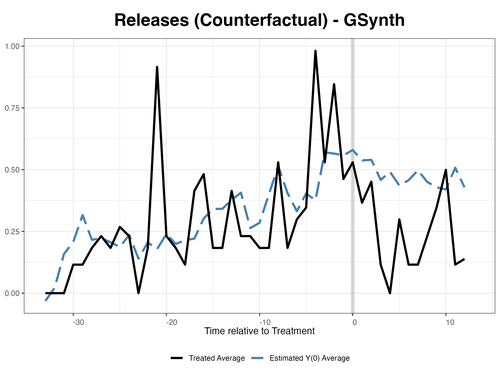}
        \caption{Releases}
    \end{subfigure}

    \vspace{0.5cm}

    \begin{subfigure}{0.48\textwidth}
        \includegraphics[width=\textwidth]{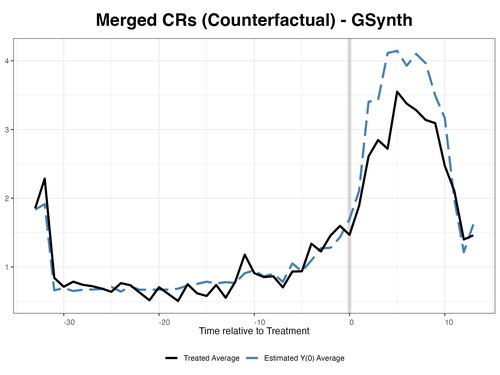}
        \caption{Merged CRs}
    \end{subfigure}
    \hfill
    \begin{subfigure}{0.48\textwidth}
        \includegraphics[width=\textwidth]{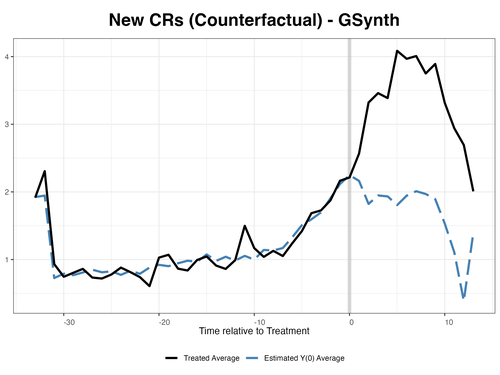}
        \caption{New CRs}
    \end{subfigure}
    \vspace{0.5cm}
    \begin{subfigure}{0.48\textwidth}
        \includegraphics[width=\textwidth]{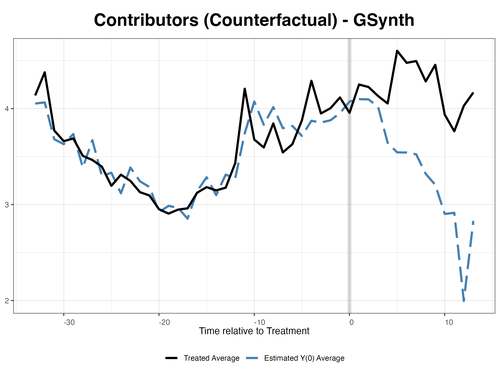}
        \caption{Contributors}
    \end{subfigure}
    \hfill
    \begin{subfigure}{0.48\textwidth}
        \includegraphics[width=\textwidth]{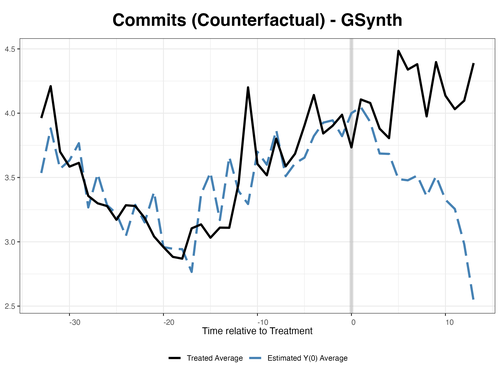}
        \caption{Commits}
    \end{subfigure}
    \caption{GSCM Counterfactuals for ATT (1 of 2)}

\end{figure}
\begin{figure}[htbp]
    \ContinuedFloat 
    \centering

    \begin{subfigure}{0.48\textwidth}
        \includegraphics[width=\textwidth]{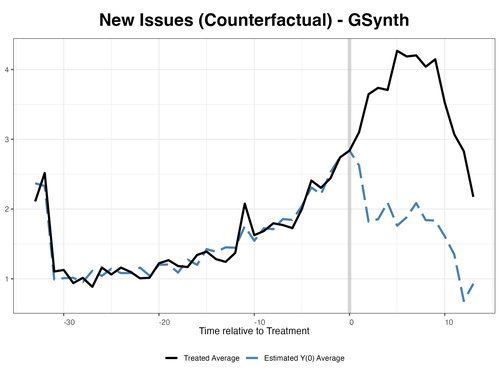}
        \caption{GSCM Counterfactuals: New Issues}
    \end{subfigure}
    \caption{GSCM Counterfactuals for ATT (2 of 2)}
    \label{fig:gsynth_counterfactuals}
\end{figure}

\begin{figure}[htbp]
    \centering
    \begin{subfigure}{0.48\textwidth}
        \includegraphics[width=\textwidth]{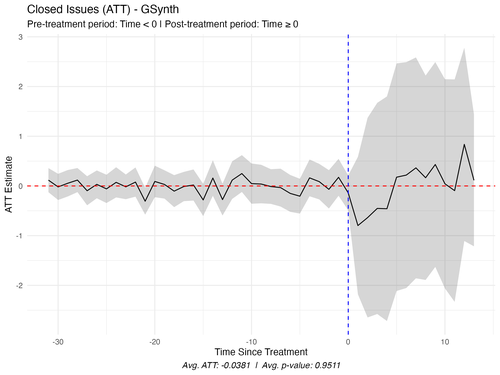}
        \caption{GSynth ATT: Closed Issues}
    \end{subfigure}
    \hfill
    \begin{subfigure}{0.48\textwidth}
        \includegraphics[width=\textwidth]{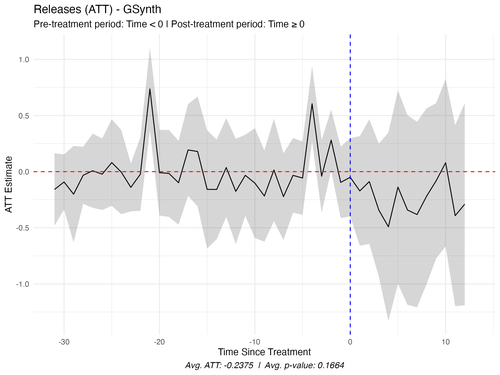}
        \caption{GSynth ATT: Releases}
    \end{subfigure}

    \vspace{0.3cm}

    \begin{subfigure}{0.48\textwidth}
        \includegraphics[width=\textwidth]{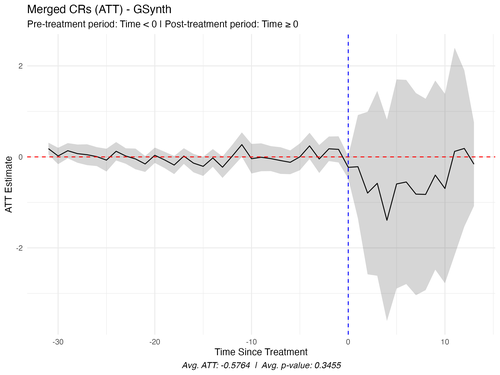}
        \caption{GSynth ATT: Merged CRs}
    \end{subfigure}
    \hfill
    \begin{subfigure}{0.48\textwidth}
        \includegraphics[width=\textwidth]{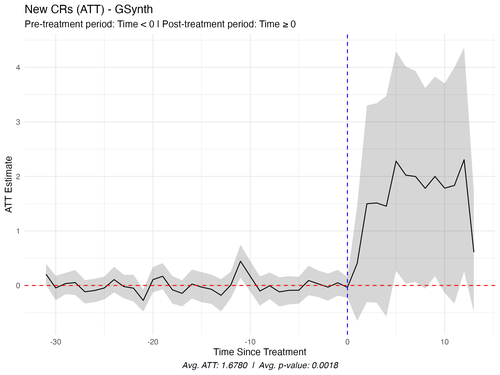}
        \caption{GSynth ATT: New CRs}
    \end{subfigure}

    \vspace{0.3cm}

    \begin{subfigure}{0.48\textwidth}
        \includegraphics[width=\textwidth]{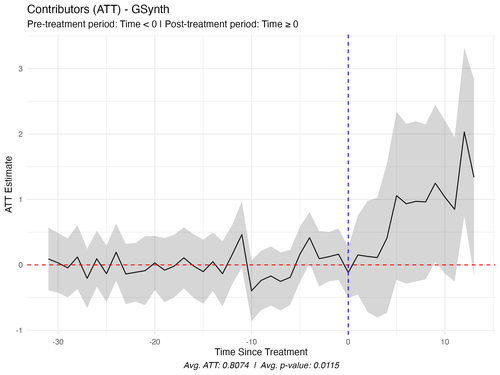}
        \caption{GSynth ATT: Contributors}
    \end{subfigure}
    \hfill
    \begin{subfigure}{0.48\textwidth}
        \includegraphics[width=\textwidth]{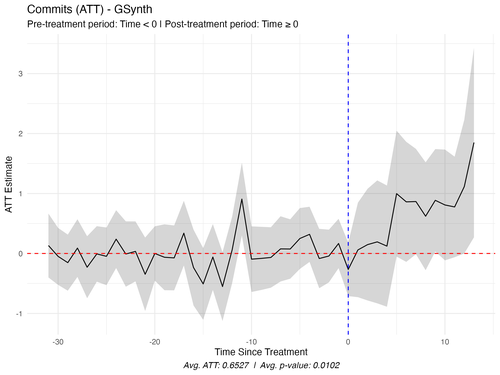}
        \caption{GSynth ATT: Commits}
    \end{subfigure}

    \caption{Generalized Synthetic Control ATT Estimates (1 of 2)}
\end{figure}

\begin{figure}[htbp]
    \ContinuedFloat 
    \centering
    \begin{subfigure}{0.48\textwidth}
        \includegraphics[width=\textwidth]{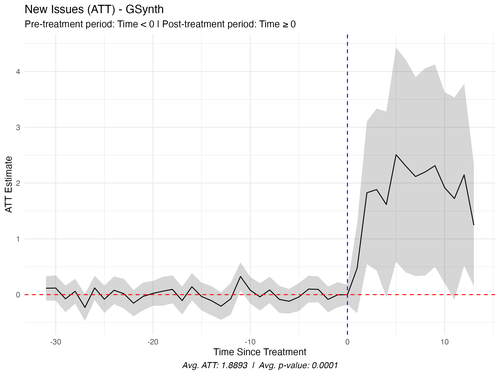}
        \caption{GSynth ATT: New Issues}
    \end{subfigure}
    \caption{Generalized Synthetic Control ATT Estimates (2 of 2)}
    \label{fig:gsynth_att}
\end{figure}

\begin{figure}[htbp]
    \centering
    \begin{subfigure}{0.48\textwidth}
        \includegraphics[width=\textwidth]{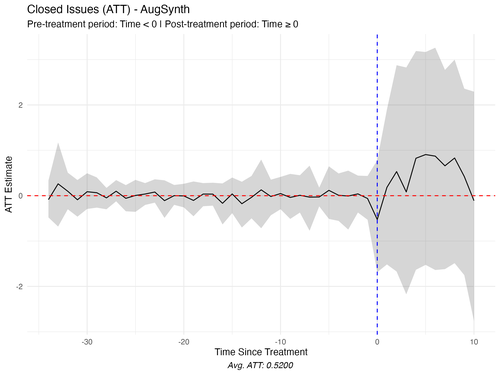}
        \caption{AugSynth ATT: Closed Issues}
    \end{subfigure}
    \hfill
    \begin{subfigure}{0.48\textwidth}
        \includegraphics[width=\textwidth]{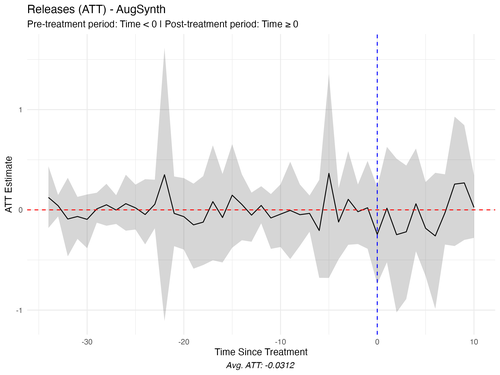}
        \caption{AugSynth ATT: Releases}
    \end{subfigure}

    \vspace{0.5cm}

    \begin{subfigure}{0.48\textwidth}
        \includegraphics[width=\textwidth]{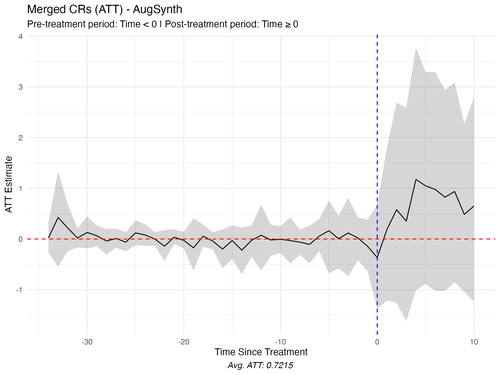}
        \caption{AugSynth ATT: Merged CRs}
    \end{subfigure}
    \hfill
    \begin{subfigure}{0.48\textwidth}
        \includegraphics[width=\textwidth]{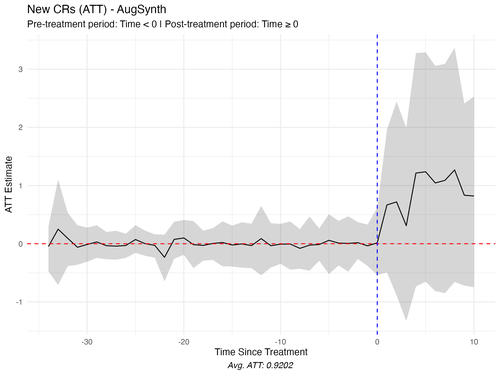}
        \caption{AugSynth ATT: New CRs}
    \end{subfigure}
    \vspace{0.5cm}
    \begin{subfigure}{0.48\textwidth}
        \includegraphics[width=\textwidth]{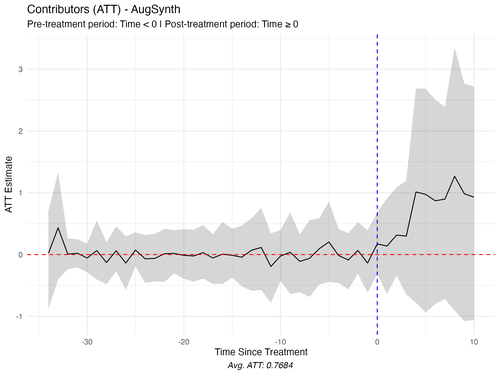}
        \caption{AugSynth ATT: Contributors}
    \end{subfigure}
    \hfill
    \begin{subfigure}{0.48\textwidth}
        \includegraphics[width=\textwidth]{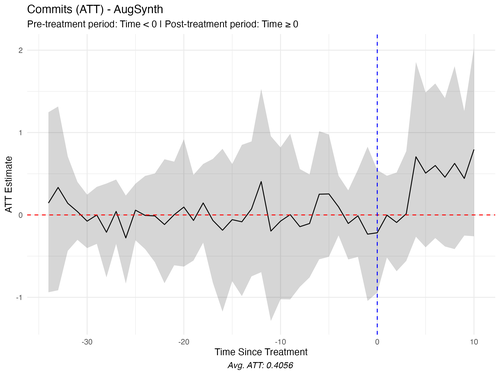}
        \caption{AugSynth ATT: Commits}
    \end{subfigure}
    \caption{Augmented Synthetic Control ATT Estimates (1 of 2)}
\end{figure}

\begin{figure}[htbp]
    \ContinuedFloat 
    \centering
    \begin{subfigure}{0.48\textwidth}
        \includegraphics[width=\textwidth]{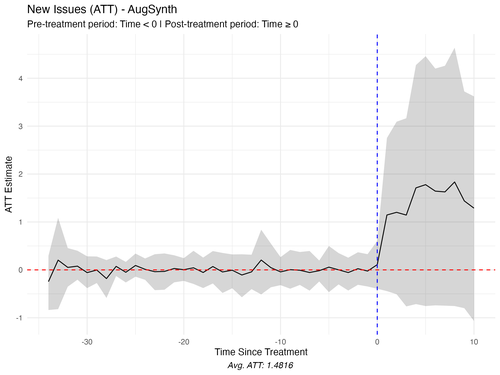}
        \caption{AugSynth ATT: New Issues}
    \end{subfigure}
    \caption{Augmented Synthetic Control ATT Estimates (2 of 2)}
    \label{fig:agusynth_att}
\end{figure}
\FloatBarrier
\subsection{Tables}

\begin{table}[h]
\centering
\caption{Model Factors and Performance Metrics}
\label{tab:model_factors}
\begin{tabular}{lcccc}
\hline
\textbf{Model} & \textbf{GSCM Factors ($r^*$)} & \textbf{MSPE} & \textbf{ASCM Factors} \\ \hline
Closed Issues   & 3  & 0.38324 & 2 \\
Releases        & 3  & 0.26368 & 4 \\
Merged CRs      & 3  & 0.28260 & 3 \\
New CRs         & 6  & 0.27553 & 4 \\
Contributors      & 7  & 0.62329 & 4 \\
Commits         & 6  & 0.64291 & 5 \\
New Issues      & 4  & 0.30165 & 4 \\ \hline
\end{tabular}
\end{table}

\subsection{Code Repository}

All materials required to replicate the analyses presented in this thesis are 
publicly available at:

\begin{center}
    \url{https://github.com/ldmnch/thesis_mds_2026}
\end{center}

\bibliographystyle{unsrt}  
\bibliography{thesis_refs}

\end{document}